\documentstyle[12pt]{article}
\begin{document}
\title{Why Are Renormalons Always In The Same Place ?}
\author{
David T. Barclay \\ 
Dept. of Mathematical Sciences \\
University of Liverpool \\
Liverpool L69 3BX, U.K.}
\date{September 1996 } 
\maketitle
\begin{abstract}
In order that current and future renormalon 
results in QCD
can be used to their full advantage it is 
important to
understand how the Borel transforms of 
related functions
are themselves related. For example, a change of
renormalisation scheme can involve the inversion 
of a function;
how is the transform of a function related to the 
transform of an inverse ?
We outline a method of extracting the dominant 
behaviour of
the Borel transform after the action of simple 
operations on the function.
\end{abstract}
\newpage
\section{Introduction} 
Borel transforms provide a convenient and
insightful method of
describing the large-order behaviour of QCD 
perturbation series.
An observable
\begin{equation}
\label{series}
R(a) = \sum_{n=0}^{\infty} r_n a^{n+1}, 
\qquad a \equiv \alpha_s /\pi
\end{equation}
is represented as the integral
\begin{equation}
\label{transform}
R(a) = \int_0^{\infty} F(z) e^{-z/a} dz
\end{equation}
where $F(z)$ is the transform.
(If necessary, principal values are used 
to define the integral --
see Appendix).
A singularity of the form
\begin{equation}
\label{renorm}
F(z) \sim {A \over (1- \alpha z)^{\beta}}, 
\qquad z \sim 1/\alpha
\end{equation}
gives rise to a factorial divergence 
\begin{equation}
\label{fac}
r_n \sim n ! A {\alpha}^n {\beta (\beta+1)
 \ldots (\beta+n-1)
				\over n!}
\end{equation}
in the perturbation expansion of $R(a)$.

Much attention has been devoted to characterising 
the $F(z)$ of
QCD observables, both in general and 
for particular cases.
Arguments related to power corrections show that
a typical observable has an 
infinite chain of singularities
of the form (\ref{renorm}), equally 
spaced along the positive 
real axis of the Borel 
$z$-plane \cite{poles}\cite{mueller}.
These are called IR renormalons 
and are crucial is assessing
the practical reliability of perturbation theory.
Although the positions and general 
form of the renormalons are
fixed for any $R(a)$ by these 
arguments, the particular details
(especially the $A$ and 
$\beta$ of a singularity) have to be 
established for each observable separately.
Such work has both reproduced 
the predicted general features 
and provided specific information 
about $F(z)$ in cases of
interest, see e.g. \cite{broadhurst}-\cite{stein}.
A notable feature of these results is that 
$F(z)$ is invariably
a simpler function than $R(a)$.

However to fully exploit the 
potential of these results it is 
often necessary to consider 
functions related to $R(a)$ rather
than the calculated $R(a)$ itself.
For example, $R(a)$ 
and its Borel transform may have been 
calculated using the $\overline{MS}$
renormalisation scheme, 
but it is now the corresponding
results in the effective 
charge scheme \cite{EC} that are 
wanted.
The function of interest is now 
the effective charge 
$\beta$-function
\begin{equation}
\label{beta}
\rho(R)  \equiv  {\beta(a(R)) \over da/dR} 
\end{equation}
where $\beta(a)$ is the 
$\overline{MS}$ $\beta$-function. 
Calculating this involves doing 
a series of operations to the 
function $R(a)$: inverting it 
to give $a(R)$, finding the 
function-of-a-function $\beta(a(R))$, 
differentiation and
division. 
In terms of the function 
these are straightforward enough
operations, though possibly difficult 
to do exactly in practice.
A natural question -- particularly 
if, as noted, the observable
is simpler as a transform than 
as a function of $a$ -- is what
effect these operations have in the Borel plane ?
If we know the Borel transform 
of $R(a)$, what does this tell
us about the transform of $a(R)$ ?

Because such operations are 
commonplace, questions like this
arise continually when 
applying the results of renormalon
calculations.
It is this practical need 
that the method outlined here 
seeks to meet.

\section{The Operations}

The operations that this paper will discuss are
\begin{enumerate}
\item  Multiplication $R_1(a) R_2 (a)$
\item  Differentiation $dR/da$
\item  Function-of-Function $R_1 (R_2 (a))$
\item  Division $R_1 (a)/ R_2 (a)$
\item  Inversion $a(R)$
\end{enumerate}
Some of these are easy and some are hard.
The first two are the straightforward ones.
Like with any simple integral 
transform, the Borel transform
of a product is given by a 
convolution integral over the 
transforms of the factors.
If $R_1 (a)$ and $R_2 (a)$ 
have transforms $F_1 (z)$ and $F_2 (z)$,
then $R_1 (a) R_2 (a)$ has a transform
\begin{equation}
\label{conv}
 F_3 (z) = \int_0^z F_1 (u) F_2 (z-u) du.
\end{equation}
This will be true even if 
the transforms contain renormalon
singularities (see Appendix).
As noted by 't Hooft \cite{thooft}, 
the presence of such singularities
in the factors leads to similar 
singularities in the product as
a result of (\ref{conv}).
This aspect is discussed in more detail below.
Indeed much of this paper can be 
considered a generalization of this
observation about the convolution integral.

The other straightforward 
operation is differentiation, since the 
transform of $a^n d^n R/ d a^n$ is
\begin{equation}
\label{diff}
{d \over dz} \biggl( z^n {d^{n-1} F 
\over d z^{n-1}}\biggr).
\end{equation}
It is possible to remove 
powers of $a$ from $a^n d^n R/da^n$ by
using
\begin{equation}
\label{arg}
 {R(a) \over a} = 
F(0) + \int_0^{\infty} e^{-z/a} F'(z) dz.
\end{equation}
The $F(0)$ piece here corresponds 
to a delta function term in
the transform which we will seek to avoid.
Thus (\ref{arg}) will only be 
used in this paper on functions
for which $F(0)=0$.

The remaining operations 3-5 
are all intrinsically difficult ones.
By this we mean that it 
is unlikely that closed-form results will
ever be found for 
their effect on transforms, in the way that
(\ref{conv}) and 
(\ref{diff}) exist for multiplication and 
differentiation.
Even with a simple example 
like $F(z) = 1/(1-z)$, it appears that
they can produce a 
transform that is much more complicated in 
its fine detail than the input one.
Our simple and natural 
question does not have a simple and 
natural full answer.
However the issue of 
practicality remains and for this a complete
answer to the problem is not necessary.
For the effective use of 
current and foreseeable QCD calculations
it will be sufficient to 
extract only the more important effects
that the use of these 
operations will have on the transforms.
Furthermore, by concentrating 
on the main effects, a simple and
natural picture does emerge.

The general question of the 
effect of the operations 3-5 on
Borel summable functions has 
been addressed before, in a pair
of papers by Auberson and Mennessier \cite{AM}.
Their main conclusion is 
that the functions thus produced are
also Borel summable.
This result is clearly a 
necessary starting point for our
investigation in that it 
establishes that the transforms we
will be trying to describe actually exist.
In general \cite{AM} provides 
a firm foundation of rigour that
will be relied upon 
implicitly throughout what follows.
As a consequence of 
the emphasis on extracting information
that can be used in 
practical QCD contexts, the approach
and language used here is 
however very different than in \cite{AM}.

\section{The $\lambda$-Expansion}

As a concrete example, 
consider operation 5, inverting a function
$R(a)$ to $a(R)$.

The central idea of our 
approach is to introduce the following
split in the initial function
\begin{equation}
\label{split}
 R(a) = a + \lambda \tilde R (a).
\end{equation}
The parameter $\lambda$ is a 
bookkeeping device which will be set
equal to one at the end.
Thus
\begin{equation}
\label{tilde}
\tilde R (a) =   \sum_{n=1}^{\infty} r_n a^{n+1}
\end{equation}
is simply all of $R(a)$ 
except for its lowest order term.
Of course (\ref{split}) really 
defines a function $R(a,\lambda)$
such that $R(a)=R(a,1)$.
And if we are starting 
from a QCD observable, $R(a)$ will
be renormalisation scheme 
invariant, but $R(a,\lambda \neq 1)$ will
not be. 
But these nuances are 
unproblematical and the $\lambda$-dependence
of $R(a,\lambda)$ will 
be suppressed in the notation.

It is also possible to think 
of $a(R)$ as now having a 
$\lambda$-dependence, 
though one more complicated than that in 
(\ref{split}).
Expand $a(R)$ as an expansion in $\lambda$.
\begin{equation}
\label{taylor}
 a(x) = x + \lambda 
{\partial a 
\over \partial \lambda}\bigg|_{\lambda=0}
 + {\lambda^2 \over 2} 
{\partial^2 a \over \partial \lambda^2}
	\bigg|_{\lambda=0} + \ldots
\end{equation}
The coefficients are 
functions of $x$ and can be found by 
repeatedly differentiating (\ref{split}) 
rewritten as
\begin{equation}
\label{split2}
 x= a(x) + \lambda \tilde R (a(x))
\end{equation}
with respect to $\lambda$.
Thus the first differentiation gives
\begin{equation}
\label{diff2}
0= {\partial a 
\over \partial \lambda} + \tilde R (a)
	+ \lambda {\partial 
\tilde R \over \partial a}
	{\partial a \over \partial \lambda}
\end{equation}
and so
\begin{equation}
\label{first}
{\partial a 
\over \partial \lambda}\bigg|_{\lambda=0}
	= - \tilde R (x).
\end{equation}
The first 
few terms of (\ref{taylor}) are found to be
\begin{eqnarray}
 a(x) = x & - & \lambda 
\tilde R (x) + \lambda^2 \tilde R (x)
		{d \tilde R 
\over dx} \nonumber \\
  & - & {\lambda^3 
\over 2} \biggl( {\tilde R}^2 
	{d^2 \tilde R 
\over dx^2} + 2 \tilde R
	\biggl( {d \tilde R 
\over dx} \biggr)^2 \biggr) 
	+ \ldots \label{expan}
\end{eqnarray}
This will be called 
the $\lambda$-expansion of $a(R)$.

Given that all the 
other perturbative expansions in this paper
are divergent, 
it is important to emphasize that the 
$\lambda$-expansion is usually convergent as
a series in $\lambda$.
A detailed discussion of 
this issue is deferred \cite{BM}, but
heuristic arguments can 
be offered for (\ref{expan}) being
convergent.
First, note that so 
far nothing has been assumed about the
convergence properties of (\ref{tilde}).
$\tilde R (a)$ can be as 
favourable and well-behaved a function
as required and (\ref{expan}) 
will still be derived.
As a relation between 
$\tilde R(x)$ and $a (x)$ as functions,
the structure of the $\lambda$-expansion 
is independent of their
behaviour as expansions in $x$.
In this light, there is no 
{\it a priori} reason to expect (\ref{expan})
to be especially 
problematic for the cases of interest. 
Secondly, when $\tilde R (x)$ 
is divergent, the resulting divergence
of $a(x)$ as a series in 
$x$ can be thought of as having been
absorbed into the $\tilde R(x)$ 
that appear in (\ref{expan}) to
leave a convergent expression.
Thirdly, consider the function
\begin{equation}
\label{yfun}
\lambda(y) = \lambda(y,x) 
\equiv {y \over \tilde R (x-y)}
\end{equation}
and its inverse $y(\lambda) 
= y(\lambda,x)$ for a fixed $x$.
$\tilde R (a)$ is expected to 
have one cut in the complex $a$-plane,
along the negative real axis.
Thus $\tilde R (x-y)$ is 
analytic about $y=0$ for all $x > 0$
and $\lambda(y)$ will converge for $|y| < x$.
Hence its inverse, 
$\lambda(y)$, has a non-zero radius of
convergence \cite{Bromwich}, but since 
\begin{equation}
\label{aasy}
a(x) = x - y(\lambda,x),
\end{equation}
so does (\ref{expan}).

Clearly none of these arguments 
is sufficient to prove that the
$\lambda$-expansion is convergent 
for $\lambda=1$, as required.
Pending the detailed discussion \cite{BM}, 
the issue is set aside.
For present purposes, we merely 
note that in the results presented
here the expansions actually 
have infinite radii of convergence.

Turning to the use of (\ref{expan}), 
for a given $R(a)$ and hence
$\tilde R (a)$, setting $\lambda=1$ 
in (\ref{expan}) provides
a systematic means of calculating 
$a(R)$ from $\tilde R (a)$.
It is a method of inverting functions.
Its importance as such is 
that it serves to reduce the hard 
operation of inversion to a 
sequence (albeit infinite) of easy 
ones, 
namely differentiation and multiplication.
Given the Borel transform of 
$R(a)$, (\ref{conv}) - (\ref{arg})
can be used to calculate the 
Borel transform of any particular
term in the $\lambda$-expansion 
and then these contributions
summed to obtain the 
Borel transform of $a(R)$.
\section{Calculating the Transform}
The simplest case is where 
the transform of $R(a)$ (and thus
$\tilde R (a)$) has a single 
singularity at $z=1/\alpha$ such
that
\begin{equation}
\label{input}
F(z) \sim {A \over (1-\alpha z)^{\beta}}, 
\qquad z \sim 1/\alpha.
\end{equation}
The generalization to multiple poles, 
as required for realistic
QCD examples, will
be straightforward.
Neglecting numerical factors, 
the general term in (\ref{expan}) is
\begin{equation}
\label{general}
\lambda^n {d^{q_1} 
\tilde R \over dx^{q_1}} \ldots
		{d^{q_n} 
\tilde R \over dx^{q_n}}
\end{equation}
\begin{equation}
\label{constraint}
q_1+ \ldots q_n = n-1.
\end{equation}
What is the transform of this, given 
an $\tilde R$ implied by (\ref{input}) ?
Using (\ref{diff})-(\ref{arg}) 
this is easily found in principle, but
for any particular $F(z)$ that 
accords with (\ref{input}) the
convolution integrals 
quickly become impossible to evaluate 
exactly.
It becomes necessary to follow 
only the more important features of
the transforms through the calculation.

Note that if $F(z)$ in (\ref{input}) 
has another singularity
at $z=1/\alpha$ with $\beta'=\beta-1$, 
then according to (\ref{fac})
the additional contribution to 
the factorial divergence of the
coefficient $r_n$ is suppressed 
by $1/n$ and so can be neglected
at large orders.
All singularities at $z=1/\alpha$ 
with smaller $\beta$ are similarily
sub-leading in $r_n$.
The singularity with largest 
$\beta$ at $z=1/\alpha$ will be called
the dominant singularity, 
the others sub-dominant.
It is the dominant 
singularities (possibly at different 
$z=1/\alpha$) that
will be most important in practice.
Furthermore, for the operations 1-5 
it turns out that sub-dominant
terms in the initial transform 
only give rise to sub-dominant terms
in the result.
From now on the sub-dominant 
singularities will be neglected.

The features of the 
transforms that are to be tracked through the
calculation will be 
their behaviour close to $z=0$,
the positions of the poles 
and the dominant behaviour there.
For any transform of interest, 
these can be summarised thus
\begin{equation}
\label{form}
F_i (z) \sim \left\{
 \begin{array}{cl}
	z^{m_i}, & z \sim 0 \\
	A_i 
(1-\alpha z)^{-\beta_i}, & z \sim 1/\alpha
 \end{array}
\right.
\end{equation}
for an input transform like (\ref{input}).

Note that all of these are 
also only singular at $z=1/\alpha$.
Why is this ?
Firstly, differentiation of 
$\tilde R$ and the use of (\ref{diff})
cannot cause the transform to 
become non-analytic at any other point.
Sub-dominant terms also remain sub-dominant.
Secondly, 
multiplication and the use of the convolution
integral (\ref{conv}) have much the same effect.
Consider where $F_1(z)$ 
and $F_2(z)$ in (\ref{conv}) are of the
form (\ref{form}).
How does $F_3(z)$ behave ?
For $z < 1/\alpha$ the integrand 
is finite and so is $F_3(z)$.
But as $z \rightarrow 1/\alpha$ 
the integrand begins to diverge
at both ends of its interval 
and hence so can $F_3(z)$ as
$z \rightarrow 1/\alpha$.
For $z > 1/\alpha$, 
a principal value is taken in the integral
where necessary 
(see Appendix) and $F_3(z)$ is finite.
The only point where 
$F_3(z)$ is non-analytic is $z=1/\alpha$.
This is essentially the 
observation made by 't Hooft \cite{thooft}.
Furthermore, if $F_1(z)$ 
and $F_2(z)$ are as in (\ref{form}), so
is $F_3(z)$, but with
\begin{equation}
\label{mmm}
m_3 = m_1 + m_2 + 1
\end{equation}
\begin{equation}
\label{bbb}
\beta_3 = \left\{
 \begin{array}{lll}
        -\beta_1+m_2+1 & 
\hbox{if} & -\beta_1+m_2+1<-\beta_2
		+m_1+1 \\
        -\beta_2+m_1+1 
& \hbox{if} & -\beta_2+m_1+1<-\beta_1
		+m_2+1
 \end{array}
\right.
\end{equation}
One has to know the 
behaviours near $z=0$ because when one
of the transforms 
in the integrand of (\ref{conv}) is
diverging to give 
the divergence in $F_3(z)$, the other
transform's argument 
is tending towards $z=0$.
All this 
is neglecting the sub-dominant terms. 
However, such terms in either 
$F_1(z)$ or $F_2(z)$ do not
produce dominant 
terms in $F_3(z)$ as a result of this
convolution; they are safely neglected.
Since differentiation 
and multiplication are the only
operations involved in 
finding (\ref{general}), its transform
and any intermediate ones 
involved in finding it will indeed
thus behave as (\ref{form}).

In addition, one can 
use (\ref{diff}), (\ref{mmm}) and (\ref{bbb})
to find the $A_i$, 
$\beta_i$ and $m_i$ of these transforms.
Most of the terms (\ref{general}) 
turn out not to contribute to
the dominant 
singularity in the transform of $a(R)$.
At each order in 
$\lambda$ there is one term of the form
\begin{equation}
\label{main}
\lambda^n 
{\tilde R}^{n-1} 
{d^{n-1} \tilde R \over dx^{n-1}}
\end{equation}
and it is only these 
that contribute to this singularity.
Working through the 
(tedious) details and summing the 
$\lambda$-expansion one 
finds that, for $\tilde R(a)$ given by
(\ref{input}), the dominant 
part of the transform for $a(R)$ is
\begin{equation}
\label{answer}
F(z) \sim {- \lambda A 
e^{-\lambda r_1 /\alpha} \over
		(1 - \alpha z)^{\beta} }, 
\qquad z \sim 1/\alpha,
\end{equation}
where $r_1$ is 
the one-loop coefficient of $R(a)$.
Finally, $\lambda$ is 
set equal to one.

The main effect that inversion 
has had on the transform is thus to
change the overall constant 
(the $A_i$ in (\ref{form})).
The $\beta_i$ doesn't change; 
this need not be the case for other
operations.
Most importantly, the position 
of the singularity hasn't changed.
In retrospect this is an 
obvious consequence of the non-obvious
fact that inversion can 
be reduced to multiplication and
differentiation.

Additional singularities at other 
positions do not change this basic
picture.
If (\ref{input}) is generalised to
\begin{equation}
\label{string}
F(z) \sim \sum_m {A_{m} 
\over (1- \alpha_m z)^{\beta_m}},
		\qquad z \sim 1/\alpha_m,
\end{equation}
(\ref{answer}) becomes
\begin{equation}
\label{strans}
F(z) \sim \sum_m 
{-A_{m} e^{-r_1/\alpha_m} \over
	(1-\alpha_m z)^{\beta_m}}, 
\qquad z \sim 1/\alpha_m.
\end{equation}
The additional 
complications (\ref{string}) 
introduces are all
sub-dominant in (\ref{strans}).

The idea of a split (\ref{split}) 
in one of the functions
leading to a $\lambda$-expansion 
also renders operations
3 and 4 tractable.
Indeed in these cases 
the $\lambda$-expansion appears much
more familiar.
\begin{equation}
\label{funfun}
R_1(x + \lambda 
\tilde R_2(x)) = \sum_{n=0}^{\infty}
	{\lambda^n \over n!} 
{\tilde R_2 (x)}^n 
	{d^n R_1 \over dx^n}
\end{equation}
\begin{equation}
\label{quot}
{R_1(x) \over x+\lambda \tilde R_2(x)} =
 {R_1(x) \over x} 
\sum_{n=0}^{\infty} \biggl( - 
	{\lambda \tilde R_2(x) 
\over x} \biggr)^n.
\end{equation}
However since these operations 
involve two functions, the
details and the 
results are contingent on the specifics 
of two input transforms 
and a discussion is again
deferred \cite{BM}.
A full set of results 
covers the situations one is liable
to be confronted with in practice.

A general feature is however clear.
Because all these 
operations can be reduced to multiplication
and differentiation, 
if the initial transforms have the 
renormalon structure predicted 
by QCD \cite{poles}\cite{mueller}, so do
the transforms produced by the operations.
The universality of that 
structure is further confirmed:
all QCD observables 
look the same in the Borel plane.

$$ $$

{\bf Acknowledgements}
$$ $$

Chris Maxwell is thanked for useful comments and
invaluable encouragement.

\section*{Appendix}

Although it is a standard 
theorem \cite{Hardy} that the Cauchy
product of two Borel summable 
series is also Borel summable,
the result for products 
required here is that the Borel transform
of the product is given by 
the convolution (\ref{conv}), even
when singularities are present.
Since we know of no 
proof of this in the literature,
one is outlined here.
Consider two transform integrals
\begin{equation}
\label{defs}
f(a) = \int_0^{\infty} 
F(z) e^{-z/a} dz, \qquad
g(a) = P \int_0^{\infty} G(z) e^{-z/a} dz,
\end{equation}
where $F(z)$ is analytic, 
but $G(z)$ has a singularity at $z=\eta$.
The $P$ indicates a principal value.
The convolution integral will thus be
\begin{equation}
\label{cint}
H(z) = P \int_0^z G(w)F(z-w) dw.
\end{equation}
Now
\begin{eqnarray}
h(a) & = & g(a)f(a) \label{m1} \\
     & = & \lim_{\epsilon \rightarrow 0} 
\biggl( \int_0^{\eta-\epsilon}
		G(z)e^{-z/a} dz 
+ \int_{\eta+\epsilon}^{\infty}
		G(z)e^{-z/a} dz \biggr) 
		f(a) \label{m2} \\
	& = & {\overline h}(a) 
   + \lim_{\epsilon 
\rightarrow 0} \int_0^{\eta-\epsilon}
	dz G(z) e^{-z/a} 
\int_{\eta-z-\epsilon}^{\eta-z+\epsilon}
	dw F(w) e^{-z/a}. \label{m3}
\end{eqnarray}
where
\begin{equation}
\label{m4}
{\overline h}(a) \equiv P 
\int_0^{\infty} H(z) e^{-z/a} dz.
\end{equation}
It is the final 
term in (\ref{m3}) that is at issue.
The details of whether
or not it tends to 
zero depend on the specific $G(z)$.
It suffices to consider whether
\begin{equation} 
\label{eps}
\epsilon \int_0^{\eta-\epsilon} G(z) dz
\end{equation}
vanishes as $\epsilon \rightarrow 0$.
If it does, then the 
natural generalisation of the convolution
theorem, 
namely that $h(a) = {\overline h}(a)$,
is true.

If, as if in QCD,
\begin{equation}
\label{type}
G(z) \sim  
(\eta - z)^{-\beta}, \qquad z \sim \eta,
\end{equation}
then (\ref{eps}) 
vanishes only if $\beta<2$.
However the naive principal 
value definition in (\ref{defs})
only holds for $\beta \leq 1$ anyway.
For $\beta >1$, 
one can define the transform via
\begin{equation}
\label{altdef}
g(a) = a^{-n} P 
\int_0^{\infty} {\overline G}(z) e^{-z/a}dz
\end{equation}
for some $n$ such that 
${\overline G}(z)$ has a singularity
with $\beta'=\beta-n<1$.
In this paper, the transforms 
are implicitly defined like this.
For clarity, the $a^{-n}$ 
factors are suppressed and $\beta>1$
ia allowed, but with 
care it can be arranged such that 
$\beta<1$  in all transforms.
The convolution theorem thus 
holds for all products considered
here.

The generalization to both $F$ 
and $G$ having multiple renormalon
singularities is straightforward.

Principal values have been 
used to define the Borel integrals
because the generalization 
of the theorem, (\ref{cint}) and
(\ref{m4}), is then particularly natural.
However it is more 
common in QCD to define Borel integrals
using contours 
that detour around singularities.
These versions have the 
disadvantage here that the contours
required for the 
equivalents of (\ref{cint}) and (\ref{m4})
are then not obvious.

\end{document}